\documentclass[preprint,showpacs,preprintnumbers,
amsmath,amssymb]{revtex4}

\usepackage{graphicx}
\usepackage{dcolumn}
\usepackage{bm,morefloats,color}

\usepackage{amssymb,bm}

\def\Eqref#1{Eq.~(\ref{#1})}

\def\Eq#1{\begin{equation} #1 \end{equation}}
\def\Eqr#1{\begin{eqnarray} #1 \end{eqnarray}}
\def\Eqrsubl#1#2{\begin{subequations}\label{#1}\Eqr{#2}\end{subequations}}

\newcommand{\nn}{\nonumber}
\newcommand{\pd}{\partial}

\newcommand{\bea}{\begin{eqnarray}}
\newcommand{\eea}{\end{eqnarray}}

\def\Ysp{{\rm Y}}
\def\Zsp{{\rm Z}}
\def\X5sp{{\rm X}_5}
\def\Y3sp{{\rm Y}_3}
\def\Z3sp{{\rm Z}_3}

\def\lap{{\triangle}}
\def\e{{\rm e}}

\begin{document}

\title{Dynamical angled brane}

\author{Kei-ichi Maeda}%
\affiliation{%
Department of Physics, Waseda University, Okubo 3-4-1,
Shinjuku, Tokyo 169-8555, Japan}%

\author{Kunihito Uzawa}
\affiliation{%
Department of Physics,
School of Science and Technology,
Kwansei Gakuin University, Sanda, Hyogo 669-1337, Japan.
}%

\date{\today}

\begin{abstract}
We discuss the dynamical D$p$-brane solutions describing any
number of D$p$-branes whose relative orientations are given by 
certain SU(2) rotations. These are the generalization of 
the static angled D$p$-brane solutions. 
We study the collision of dynamical D3-brane with angles in 
type II string theory, 
and show that the particular orientation of the smeared D3-brane 
configuration 
can provide an example of colliding branes if they have the same charges.
Otherwise a singularity appears before D3-branes collide.
\end{abstract}


\maketitle

\section{Introduction}
 \label{sec:in}
The time-dependent brane solutions in string theory were introduced 
in \cite{Gibbons:2005rt, Chen:2005jp, Kodama:2005fz, Kodama:2005cz, 
Kodama:2006ay, Binetruy:2007tu, Binetruy:2008ev, 
Maeda:2009zi, Gibbons:2009dr, Maeda:2009ds, Uzawa:2010zza, Maeda:2010ja, 
Minamitsuji:2010fp, Maeda:2010aj, Minamitsuji:2010kb, Uzawa:2010zz, 
Nozawa:2010zg, Minamitsuji:2010uz, Maeda:2011sh, Minamitsuji:2011jt, 
Maeda:2012xb, Minamitsuji:2012if, Blaback:2012mu, Uzawa:2013koa, 
Uzawa:2013msa, Blaback:2013taa, Uzawa:2014kka, Uzawa:2014dra, Maeda:2015joa} 
and have been widely used ever since. 
But some aspects of the physical properties such as dynamical $p$-brane 
oriented at angles have remained slightly unclear. 

The angled D$p$-brane solutions 
was discovered by \cite{Breckenridge:1997ar}, 
following the recognition of the several  
role of D-brane configurations in string theory
\cite{Hambli:1997uq, Balasubramanian:1997az, Youm:1997hw}. 
The classical solution on ten-dimensional spacetime 
that should be obtained to introduce 
oriented at angles in string theories were studied from several points of 
view in \cite{Breckenridge:1996tt, Breckenridge:1997ar, 
Youm:1997hw, Maeda:2012xb}. 
They have discussed the link between the supersymmetric configurations, 
intersection of brane and the rotation angles. 
The other completely consistent computations for angled D-branes 
were performed in \cite{Hambli:1997uq, Balasubramanian:1997az}. 
The articles \cite{Youm:1997hw, 
Di Vecchia:1999rh, Di Vecchia:1999fx} 
give a thorough review of much of what was known in the late 1990's.

These results are more transparent if the dynamical D$p$-brane 
solution in string theory is discussed in terms of supergravity theory, 
aiming to reduce everything to ordinary higher-dimensional general relativity. 
In the present paper, we will construct new solutions with nontrivial 
angles in string theory and study the dynamical behavior that one would 
expect of a string theory with the time-dependent fields and branes.

We start by describing the appropriate ansatz of fields and metric  
in string theory, and 
find the solution of dynamical 
D2-brane with angle 
in section \ref{sec:D2}. 
The solution is straightforwardly generalized from the static angled 
D2-brane background.
This is a good illustration, and also an 
important example in order to understand the orientation
of a particular type brane configuration, 
which is so called the angled D-brane system
\cite{Breckenridge:1997ar, Hambli:1997uq, Balasubramanian:1997az}.  

Using the T-duality map between the type IIA and IIB string theories 
we can obtain the dynamical angled D3- and D4-brane solutions in terms of 
dynamical D2-branes. 
In section \ref{sec:D3}, we study 
the time evolution of the dynamical 
D3-brane background in detail, and also apply this 
analysis to the D4-brane with rotation angles in sec.~\ref{sec:D4}.  
We will summarize the general  dynamical $N$ D$p$-brane system 
with angles in 
sec.~\ref{sec:g}, and  
provide more details about the collision of two D3-brane 
in the presence of angles in sec.~\ref{sec:c}.
Finally, in section \ref{sec:d} 
is devoted to summary and discussion. 

\section{Dynamical D$p$-brane system with angles}

\subsection{D2-brane with angle}
\label{sec:D2}

In this section, we consider dynamical D2-branes 
oriented at some angles in ten dimensions. 
The ten-dimensional spacetime metric depends on time as well as 
the rotation angles which describe the orientations of  various D2-branes. 
First, we write down the Einstein equations under a particular 
ansatz for fields, which is a generalization
of the metric form of known static angled D2-brane solutions. 
Then, we solve the Einstein equations and present the dynamical 
solutions explicitly. 

The action for the D2-brane system in the Einstein 
frame is written as
\Eqr{
S&=&\frac{1}{2\kappa^2}\int \left(R\ast{\bf 1}
 -\frac{1}{2}\ast d\phi \wedge d\phi
 -\frac{1}{2\cdot 4!}\e^{\phi/2}\ast F_{(4)}\wedge F_{(4)}\right),
\label{D2:action:Eq}
}
where $\kappa^2$ is the ten-dimensional gravitational constant, 
$\ast$ is the Hodge dual operator in the ten-dimensional spacetime, 
and $F_{(4)}$ is 4-form field strength. We assume that 
the 4-form $F_{(4)}$ in the action \Eqref{D2:action:Eq} is now given  
by
\Eq{
F_{(4)}=dC_{(3)},
}
where $C_{(3)}$ is the 3-form gauge field. 
After variations with respect to the metric, the scalar field, 
and the gauge field, we obtain the field equations,
\Eqrsubl{D2:equations:Eq}{
&&R_{MN}=\frac{1}{2}\pd_M\phi \pd_N \phi+\frac{1}{2\cdot 4!}\e^{\phi/2} 
\left(2F_{MABC} {F_N}^{ABC}-\frac{3}{8}g_{MN} F_{(4)}^2\right),
   \label{D2:Einstein:Eq}\\
&&d\ast d\phi=\frac{1}{4\cdot 4!}\e^{\phi/2}\ast F_{(4)}\wedge F_{(4)}\,,
   \label{D2:scalar:Eq}\\
&&d\left(\e^{\phi/2}\ast F_{(4)}\right)=0\,.
   \label{D2:F:Eq}
}

\subsubsection{Single D2-brane}
First we consider a single D2-brane solution.
The geometry of dynamical D2-brane configuration
 in ten-dimensional spacetime is assumed to be 
\Eqrsubl{D2s:metric:Eq}{
ds^2&=&h^{-5/8}(t, z)\left(-dt^2+\gamma_{ij}dy^idy^j\right)
+h^{3/8}(t, z)u_{ab}(z)dz^adz^b\,,\\
\gamma_{ij}dy^idy^j&=&\delta_{ij}dy^idy^j+f(t, z)\left[
\left(\cos\theta dy^1-\sin\theta dy^2\right)^2
+\left(\cos\theta dy^3+\sin\theta dy^4\right)^2\right],\\
h(t, z)&=&1+f(t, z)\,,
}
where $ds^2$ is the line element in the Einstein frame in ten dimensions, 
$\delta_{ij}$ is the four-dimensional flat Euclidean metric, 
and $u_{ab}$ is the metric of the five-dimensional space $\Zsp$
depending only on the five-dimensional coordinates $z^a$. 
The dilaton field and the gauge field strengths are assumed to be
\Eqrsubl{D2s:fields:Eq}{
\e^{\phi}&=&h^{1/2}\,,
  \label{D2s:phi:Eq}\\
F_{(4)}&=&d\left[1-h^{-1}(t, z)\right]
\wedge dt \wedge\left(
-\sin^2\theta dy^1\wedge dy^3
+\sin\theta\cos\theta dy^1\wedge dy^4\right.\nn\\
&&\left.-\sin\theta\cos\theta dy^2\wedge dy^3
+\cos^2\theta dy^2\wedge dy^4
\right)\,.
  \label{D2s:form:Eq}
}

We first solve the equation for the gauge field strength $F_{(4)}$\,.
Substituting the above ansatz for the fields and the metric form into Eq.~(\ref{D2:F:Eq}), 
we find
\Eq{
\pd_t\pd_a h=0\,,~~~~~\lap_\Zsp h=0\,. 
  \label{D2s:gauge:Eq}
}
Then, the function $h$ is given by  
\Eq{
h(t, z)=h_0(t)+h_1(z)\,.
  \label{D2s:h:Eq}  
}

Then we solve the equation for  the scalar field; Eq.~(\ref{D2:scalar:Eq}).
Using the assumptions (\ref{D2s:metric:Eq})
and (\ref{D2s:form:Eq}), we have  
\Eq{
h^{-3/8}\left(\pd_t^2h-h^{-1}\lap_\Zsp h\right)=0\,,
  \label{D2s:scalar2:Eq}
  }
where $\triangle_{\Zsp}$
is the Laplace operator on $\Zsp$. Combining Eq.~(\ref{D2s:scalar2:Eq}) 
with Eqs.~(\ref{D2s:gauge:Eq}), (\ref{D2s:h:Eq}), we find
\Eq{
\pd_t^2h_0=0\,,~~~~~~\lap_\Zsp h_1=0\,.
}

Finally we analyze the Einstein equations~(\ref{D2:Einstein:Eq}). 
Using the ansatz (\ref{D2s:metric:Eq}) and (\ref{D2s:fields:Eq}), 
the Einstein equations become
\Eqrsubl{D2s:cE:Eq}{
&&\frac{11}{16}h^{-1}\pd_t^2h+\frac{5}{16}h^{-2}\lap_\Zsp h=0\,,
  \label{D2s:cE-tt:Eq}\\
&&h^{-1}\pd_t\pd_a h=0\,,
  \label{D2s:cE-ta:Eq}\\
&&\frac{5}{16}h^{-1}\gamma_{ij}
\left(\pd_t^2h-h^{-1}\lap_{\Zsp}h\right)+\frac{1}{2}h^{-1}
\lap_{\Zsp}\gamma_{ij}=0\,,
  \label{D2s:cEinstein-ij:Eq}\\
&&R_{ab}(\Zsp)+\frac{3}{16}u_{ab}\left(\pd_t^2h-h^{-1}\lap_\Zsp h\right)=0\,,
  \label{D2s:E-ab:Eq}
}
where $R_{ab}(\Zsp)$ is the Ricci tensor
of the five-dimensional space $\Zsp$. 
Hence, the field equations reduce to 
\Eqrsubl{D2s:fields2:Eq}{
&&R_{ab}(\Zsp)=0\,,
   \label{D2s:Ricci:Eq}\\
&&h(t, z)=h_0(t)+h_1(z)\,,~~~~\pd_t^2h_0=0\,,~~~~\lap_{\Zsp}h_1=0\,,~~~~
\lap_{\Zsp}f=0\,.
   \label{D2s:h2:Eq}
}

This is the exact solution of the present system 
 for any given Ricci flat metric $u_{ab}$.
 
As an example, we set
\Eq{
u_{ab}=\delta_{ab}\,,
 \label{D2s:flat:Eq}
 }
where $\delta_{ab}$ is the five-dimensional Euclidean metric. 
Then we find an exact solution for a single brane solution:
\Eq{
h_0(t)=c_0t+c_1\,,~~~
h_1(z)=c_2+ \frac{Q_1}{|z^a-z^a_{(1)}|^{3}}\,,
  \label{D2s:solution:Eq}
} 
where  $c_0, c_1, c_2$, $Q_1$, and $z^a_{(1)}$
 are constants.  $Q_1$ is the charge of the D2-brane,
  and $z^a_{(1)}$ is the position of D2-brane.
  The mass of the brane is given by $M_1=|Q_1|$.
If, however, some spatial dimensions of $\Zsp$ space are smeared by the D2-branes,
we will find 
 \Eq{
h_0(t)=c_0t+c_1\,,~~~
h_1(z)=c_2+ \frac{Q_1}{|z^a-z^a_{(1)}|^{3-d_\Zsp}}\,,
  \label{D2s:solution2:Eq}
} 
where $d_\Zsp (<5)$ is the smeared dimensions.
For the case of $d_\Zsp=3$, we have to replace $h_1$ with 
 \Eq{
h_1(z)=c_2+ Q_1\ln {|z^a-z^a_{(1)}|^{3-d_\Zsp}}\,.
}

  Although the above exact solution includes the orientation angle $\theta$ in the solution, 
  it is equivalent to a single brane solution without an angle.
  In fact we find that the above solution is reduced to a single D2-brane solution without an angle 
  if we rotate the $y_1$-$y_2$, and $y_3$-$y_4$ planes by the angles $\theta$ and $-\theta$, 
  respectively, as
 \begin{eqnarray}
 \left[ 
\begin{array}{c}
 {y^1}'\\
 {y^2}'\\
\end{array} 
\right]
=
 \left( 
 \begin{array}{cc}
\cos\theta&-\sin\theta\\
\sin\theta&\cos\theta\\
\end{array} 
\right)
\left[ 
\begin{array}{c}
 {y^1}\\
 {y^2}\\
\end{array} 
\right]
\,,~~~
\left[ 
\begin{array}{c}
 {y^3}'\\
 {y^4}'\\
\end{array} 
\right]
=
 \left( 
 \begin{array}{cc}
\cos\theta&\sin\theta\\
-\sin\theta&\cos\theta\\
\end{array} 
\right)
\left[ 
\begin{array}{c}
 {y^3}\\
 {y^4}\\
\end{array} 
\right]
\,.
 \end{eqnarray}
 
 Hence 
 if every $N$ D2-branes are parallel each other, we find the exact solution for $N$ D2-branes with 
 the same angle $\theta$ 
 as Eq. (\ref{D2s:metric:Eq}) with Eq. (\ref{D2s:h:Eq}) and
 \Eq{
h_0(t)=c_0t+c_1\,,~~~
h_1(z)=c_2+\sum_{\alpha=1}^N \frac{Q_\alpha}{|z^a-z^a_{(\alpha)}|^{3-d_\Zsp}}\,,
  \label{D2s:solution3:Eq}
} 
where $c_0, c_1, c_2$, $Q_\alpha$, and  $z^a_{(\alpha)}$ 
are constants and  $d_\Zsp  (=0,\cdots, 4) $ is the smeared dimensions.  
 In the case of $d_\Zsp=3 $,
  the power function must be replaced by $\ln |z^a-z^a_{(\alpha)}|$.

$Q_\alpha$, and  $z^a_{(\alpha)}$ denote the charges of branes and their positions, respectively.
This solution is the same as that without an angle.

\subsubsection{Two D2-Branes}
Next we construct the exact solutions of two intersecting D2-branes  with non-trivial angle\,. 
The ten-dimensional metric, scalar field and a single 3-form potential are 
assumed to be 
\Eqrsubl{D2t:fields:Eq}{
ds^2&=&h^{-5/8}(t, z)\left(-dt^2+\gamma_{ij}dy^idy^j\right)
+h^{3/8}(t, z)u_{ab}(z)dz^adz^b\,,\\
\gamma_{ij}dy^idy^j&=&\delta_{ij}dy^idy^j+
f_{(1)}\left[\left(\cos\theta_1 dy^1-\sin\theta_1 dy^2\right)^2
+\left(\cos\theta_1 dy^3+\sin\theta_1 dy^4\right)^2\right]\nn\\
&&~~~+f_{(2)}\left[\left(\cos\theta_2 dy^1-\sin\theta_2 dy^2\right)^2
+\left(\cos\theta_2 dy^3+\sin\theta_2 dy^4\right)^2\right]\,,
   \label{D2t:metric:Eq}\\
h(t, z)&=&1+f_{(1)}(t, z)+f_{(2)}(t, z)+f_{(1)}(t, z)f_{(2)}(t, z)\sin^2\theta\,,
  \label{D2t:h:Eq}\\
\e^{\phi}&=&h^{1/2}\,,
  \label{D2t:phi:Eq}\\
C_{(3)}&=&h^{-1}(t, z)
 dt \wedge\left[-\left\{f_{(1)}\sin^2\theta_1+f_{(2)}\sin^2\theta_2
 +f_{(1)}f_{(2)}\sin^2\left(\theta_1-\theta_2\right)\right\}
 dy^1\wedge dy^3\right.\nn\\
&&+\left(f_{(1)}\sin\theta_1\cos\theta_1 
+f_{(2)}\sin\theta_2\cos\theta_2 \right)
\left(dy^1\wedge dy^4-dy^2\wedge dy^3\right)\nn\\
&&\left.+\left\{f_{(1)}\cos^2\theta_1+f_{(2)}\cos^2\theta_2
 +f_{(1)}f_{(2)}\cos^2\left(\theta_1-\theta_2\right)\right\}dy^2\wedge dy^4
\right]\,,
  \label{D2t:form:Eq}
}
where $\theta=\theta_2-\theta_1$. 
The notable property with the non-trivial angle appears 
in the non-linear interaction term in Eq.  (\ref{D2t:h:Eq}).

In the following, we consider two D2-brane system with the different rotation angles $(\theta_1, \theta_2)$
where $\theta_1\neq\theta_2$.  We assume $(\theta_1, \theta_2)=(0,\theta)~(\theta\neq 0)$ 
without loss of generality,  
because we can always set one angle to be zero by an appropriate rotation of $y^i$-coordinates.

In terms of the ansatz for fields and metric, the gauge field equation 
(\ref{D2:F:Eq}) gives
\Eq{
\pd_t\pd_a f_{(1)}=0\,,~~~~~\lap_\Zsp f_{(1)}=0\,,~~~~~
\pd_t\pd_a f_{(2)}=0\,,~~~~~\lap_\Zsp f_{(2)}=0\,. 
  \label{D2t:gauge:Eq}
} 
From the equation of scalar field, we find 
\Eq{
h^{-3/8}\left[\pd_t^2h-h^{-1}\left\{
\left(1+f_{(2)}\sin^2\theta \right)\lap_\Zsp f_{(1)}
+\left(1+f_{(1)}\sin^2\theta \right)\lap_\Zsp f_{(2)}
\right\}\right]=0\,.
  \label{D2t:scalar2:Eq}
  }
Under our ansatz, the Einstein equations become 
\Eqrsubl{D2t:cE:Eq}{
&&\hspace{-1cm}\frac{11}{16}h^{-1}\pd_t^2h+\frac{5}{16}h^{-2}
\left[
\left(1+f_{(2)}\sin^2\theta \right)\lap_\Zsp f_{(1)}
+\left(1+f_{(1)}\sin^2\theta \right)\lap_\Zsp f_{(2)}
\right]=0\,,
  \label{D2t:cE-tt:Eq}\\
&&\hspace{-1cm}h^{-1} \left[\left(1+f_2\sin^2\theta\right)\pd_t\pd_a f_1
+\left(1+f_{(1)}\sin^2\theta\right)\pd_t\pd_a f_{(2)}\right]=0\,,
  \label{D2t:cE-ta:Eq}\\
&&\hspace{-1cm}\frac{5}{16}h^{-1}\gamma_{ij}
\left[\pd_t^2h-h^{-1}\left\{
\left(1+f_{(2)}\sin^2\theta \right)\lap_\Zsp f_{(1)}
+\left(1+f_{(1)}\sin^2\theta \right)\lap_\Zsp f_{(2)}
\right\}\right]
+\frac{1}{2}h^{-1}
\lap_{\Zsp}\gamma_{ij}=0\,,
  \label{D2t:cEinstein-ij:Eq}\\
&&\hspace{-1cm}R_{ab}(\Zsp)+\frac{3}{16}u_{ab}\left[\pd_t^2h-h^{-1}\left\{
\left(1+f_{(2)}\sin^2\theta \right)\lap_\Zsp f_{(1)}
+\left(1+f_{(1)}\sin^2\theta \right)\lap_\Zsp f_{(2)}
\right\}\right]=0\,,
  \label{D2t:E-ab:Eq}
}
where $R_{ab}(\Zsp)$ is the Ricci tensor of the $\Zsp$ space.
Then, the field equations reduce to 
\Eqrsubl{D2t:eq:Eq}{
&&R_{ab}(\Zsp)=0\,,
   \label{D2t:Ricci:Eq}\\
&&f_{(1)}(t, z)=\tilde{f}_{(1)}(t)+\bar{f}_{(1)}(z)\,,~~~~
f_{(2)}(t, z)=\tilde{f}_{(2)}(t)+\bar{f}_{(2)}(z)\,,
   \label{D2t:h2:Eq}\\
&&\pd_t^2\tilde{f}_{(1)}=0\,,~~~~\pd_t^2\tilde{f}_{(2)}=0\,,~~~~
\pd_t\tilde{f}_{(1)}\pd_t\tilde{f}_{(2)}=0\,,~~~~
\lap_{\Zsp}\bar{f}_{(1)}=0\,,~~~~\lap_{\Zsp}\bar{f}_{(2)}=0\,.
   \label{D2t:h3:Eq}
   }
For a given Ricci flat $\Zsp$ space, we can obtain the exact solution for two D2-brane system with different angles
 by solving the above equation (\ref{D2t:h3:Eq}).
 As a result, at least one of $\tilde f_{(1)}$ and $\tilde f_{(2)}$ must be constant and then
  only one brane can be time-dependent, i.e., for $\alpha=1$ or 2, 
   \begin{eqnarray}
 \tilde{f}_{(\alpha)}=c_{(\alpha)0}\,t+c_{(\alpha)1}
 \,,
 \end{eqnarray}
 where $c_{(\alpha) 0}, c_{(\alpha)1}$ are constant, 
 and 
 the other $\tilde{f}_{(\beta)}=c_{(\beta)1}$ ($\beta\neq \alpha$) are constant.

\subsubsection{Generalization to $N$  D2-brane system with different angles}
Following the same procedure as the case of two D2-brane, 
we can generalize the solution found in the
previous section to the $N$  D2-brane system with different angles 
$\theta_\alpha (\alpha=1, \cdots, N)$.
. 

The ten-dimensional metric \cite{Breckenridge:1997ar, Youm:1997hw} 
is given by 
\Eqrsubl{D2:metric:Eq}{
ds^2&=&h^{-5/8}\left[-dt^2+\gamma_{ij}dy^idy^j
+h u_{ab}(\Zsp)dz^adz^b\right],\\
\gamma_{ij}dy^idy^j&=&\delta_{ij}dy^idy^j+
\sum_{\alpha=1}^Nf_{(\alpha)}\left[
\left\{{\left(R_{(\alpha)}\right)^1}_idy^i\right\}^2
+\left\{{\left(R_{(\alpha)}\right)^3}_jdy^j\right\}^2\right],\\
h(t, z)&=&1+f(t, z)\,.
}
The metric (\ref{D2:metric:Eq}) denotes $N$ D2-brane system such that each brane first lying 
in the $(y^2, y^4)$ plane rotates  by an angle $\theta_{\alpha}$ ($\alpha=1, \cdots, N$) in the 
$(y^1, y^2)$ and $(y^3, y^4)$ planes  as 
$(\xi, \eta)\rightarrow (\e^{i\theta_{\alpha}}\xi, \e^{-i\theta_{\alpha}}\eta)$ 
where $\xi=y^1+iy^2$ and $\eta=y^3+iy^4$,
which rotation belongs to SU(2) group.
The function $h$ and the rotation matrix $R_{(\alpha)}$ associated 
with the $\alpha$th D2-brane is given by
\Eqrsubl{D2:fields:Eq}{
f(t, z)&=&\sum_{\alpha=1}^Nf_{(\alpha)}+\sum_{\alpha<\beta}^N
f_{(\alpha)}f_{(\beta)}\sin^2\left(\theta_{\alpha}-\theta_{\beta}\right),
\label{D2:f:Eq}\\
R_{(\alpha)}&=&\left(\begin{array}{cc}
	{\begin{array}[t]{cr} \cos \theta_{\alpha} & -\sin \theta_{\alpha} \\
				\sin \theta_{\alpha} &\cos \theta_{\alpha}
\end{array}} &
				{\raisebox{-10pt}{\rm\huge 0}}\\
				{\raisebox{-6pt}{\rm\huge 0}} &
{\begin{array}{rc}
				\cos \theta_{\alpha} & \sin \theta_{\alpha} \\
				- \sin \theta_{\alpha} & \cos \theta_{\alpha}
\end{array}}
	\end{array} \right)
}
where $R_{(\alpha)} (\alpha=1,\cdots, N)$ are SO(4) ($\cong$ SU(2)) matrices that 
correspond to the rotation of D2-branes. 
The scalar field $\phi$ and 
the 3-form gauge field $C_{(3)}$ are given by
\Eqrsubl{D2:solution:Eq}{
C_{(3)}&=&h^{-1}dt\wedge \left\{\sum_{\alpha=1}^Nf_{(\alpha)}
\left(R_{(\alpha)}\right)^2_idy^i\wedge 
\left(R_{(\alpha)}\right)^4_jdy^j\right.\nn\\
&&\left. ~~~~~~~-\sum_{a<b}^N f_{(\alpha)}f_{(\beta)}
\sin^2\left(\theta_{\alpha}-\theta_{\beta}\right)
\left(dy^1\wedge dy^3-dy^2\wedge dy^4\right)\right\},\\
\e^{2\phi}&=&h^{1/2}. 
}
The assumption for fields are a straightforward generalization of
the case of a static D2-brane system with certain SU(2) angle, 
in the type IIA low energy effective string theory 
\cite{Breckenridge:1997ar, Youm:1997hw}. 
From the field equations, 
the ten-dimensional metric (\ref{D2:metric:Eq}) have to obey
\Eqrsubl{D2:eq:Eq}{
&&R_{ab}(\Zsp)=0\,,
   \label{D2:Ricci:Eq}\\
&&f_{(\alpha)}(t, z)=\tilde{f}_{(\alpha)}(t)+\bar{f}_{(\alpha)}(z)\,,
   \label{D2:h:Eq}\\
&&\pd_t^2\tilde{f}_{(\alpha)}=0\,,~~~~
\pd_t\tilde{f}_{(\alpha)}\pd_t \tilde{f}_{(\beta)}=0\,,~(\alpha\ne\beta)~~~~
\lap_{\Zsp}\bar{f}_{(\alpha)}=0\,.
   \label{D2:h2:Eq}
} 

This gives the exact solution for $N$ D2-brane system with different angles
for giving  Ricci flat $\Zsp$ space.
 As two brane system,  only one brane can be time-dependent, 
 \begin{eqnarray}
 \tilde{f}_{(\alpha)}=c_{(\alpha) 0} t+c_{(\alpha)1}
 \,,
 \end{eqnarray}
 where $c_{(\alpha) 0}, c_{(\alpha)1}$ are constant,
 and 
 the other $\tilde{f}_{(\beta)}=c_{(\beta)1}$ ($\beta\neq \alpha$) are constant.
The interesting property for the case with the non-trivial angle is 
 the non-linear interaction term found in Eq.  (\ref{D2:f:Eq}).

For the case $u_{ab}=\delta_{ab}$ in more detail, where $\delta_{ab}$ are 
the five-dimensional flat Euclidean metric. 
The functions $\bar{f}_{(\alpha)}$ are harmonic 
which are associated with D2-branes located at
$z^a=z^a_{(\alpha)}$: 
\Eq{
\bar{f}_{(\alpha)}(z)=\bar c_{\alpha}+\frac{Q_{\alpha}}{|z^a-z^a_{(\alpha)}|^{3- d_\Zsp}},
}
where $c_{\alpha}$, $Q_{\alpha}$ and $z^a_{(\alpha)}$ are constants and 
  $d_\Zsp  (=0,\cdots, 4) $ is the smeared dimensions in the $\Zsp$ space.
   In the case of $d_\Zsp=3 $,
  the power function must be replaced by $\ln |z^a-z^a_{(\alpha)}|$.

\subsection{D3-brane with angle}
\label{sec:D3}
Next, we study D3-brane system with angle, which is obtained 
by a T-duality transformation from  the solutions 
presented in Sec.~\ref{sec:D2}. 
Let us consider the D2-brane solutions (\ref{D2:metric:Eq})\,.  
To perform a T-duality along the $z^5$ direction, we have to 
smear and delocalize the solution in this direction\,. 
Hence $h$ does not depend on $z^5$\,. 

The ten-dimensional T-duality map from the type IIA theory to type IIB 
theory is given by \cite{Bergshoeff:1995as, Breckenridge:1997ar}
\Eqr{
&&g^{(\rm B)}_{\zeta\zeta}=\frac{1}{g^{(\rm A)}_{\zeta\zeta}}\,,~~~~
g^{(\rm B)}_{MN}=g^{(\rm A)}_{MN}
-\frac{g^{(\rm A)}_{\zeta M}g^{(\rm A)}_{\zeta N}
-B^{(\rm A)}_{\zeta M}B^{(\rm A)}_{\zeta N}}
{g^{(\rm A)}_{\zeta\zeta}}\,,~~~~
g^{(\rm B)}_{\zeta M}=-\frac{B^{(\rm A)}_{\zeta M}}{g^{(\rm A)}_{\zeta\zeta}}\,,\nn\\
&&\e^{2\phi_{(\rm B)}}=\frac{\e^{2\phi_{(\rm A)}}}{g^{(\rm A)}_{\zeta\zeta}}\,,~~~~
B^{(\rm B)}_{MN}=B^{(\rm A)}_{MN}
+2\frac{g^{(\rm A)}_{\zeta[M}\,B^{(\rm A)}_{N]\zeta}}{g^{(\rm A)}_{\zeta\zeta}},~~~~
B^{(\rm B)}_{\zeta M}=-\frac{g^{(\rm A)}_{\zeta M}}{g^{(\rm A)}_{\zeta\zeta}},\nn\\
&&C_{MN}=C_{MN\zeta}-2C_{[M}B_{N]\zeta}^{(\rm A)}
+2\frac{g^{(\rm A)}_{\zeta[M}\,B^{(\rm A)}_{N]\zeta}C_z}{g^{(\rm A)}_{\zeta\zeta}}\,,
~~~~C_{\zeta M}=C_{M}
-\frac{C^{(\rm A)}_{\zeta}\,g^{(\rm A)}_{\zeta M}}{g^{(\rm A)}_{\zeta\zeta}}\,,\nn\\
&&C_{MNP\zeta}=C_{MNP}-\frac{3}{2}\left(
C_{[M}\,B^{(\rm A)}_{NP]}
-\frac{g^{(\rm A)}_{\zeta[M}B^{(\rm A)}_{NP]}\,C_\zeta}
{g^{(\rm A)}_{\zeta\zeta}}
+\frac{g^{(\rm A)}_{\zeta[M}\,C_{NP]\zeta}}
{g^{(\rm A)}_{\zeta\zeta}}\right)\,,~~~~C_{(0)}=-C_\zeta\,,
   \label{D3:T:Eq}
}
where $\zeta=z^5$ is the coordinate to which the T dualization is applied, and 
$M$, $N$, $P$ denote the other coordinates; $0, y^i (i=1,\cdots,4)$, and $z^a (a=1,\cdots,4)$.
Via the T-duality map \eqref{D3:T:Eq}, 
we obtain the metric of dynamical $N$ D3-brane system in ten-dimensional background 
as \cite{Breckenridge:1997ar, Youm:1997hw} 
\Eqrsubl{D3:metric:Eq}{
ds^2&=&h^{-1/2}\left[-dt^2+\gamma_{ij}dy^idy^j+d\zeta^2
+h u_{ab}(z)dz^adz^b\right],\\
\gamma_{ij}dy^idy^j&=&\delta_{ij}dy^idy^j+
\sum_{\alpha=1}^Nf_{(\alpha)}\left[
\left\{{\left(R_{(\alpha)}\right)^1}_idy^i\right\}^2
+\left\{{\left(R_{(\alpha)}\right)^3}_jdy^j\right\}^2\right],\\
h(t, z)&=&1+f(t, z)\,.
}
where 
the function $f$ and the rotation matrix $R_{(\alpha)}$ associated 
with the $\alpha$th D3-brane are given by (\ref{D2:fields:Eq})\,.

This describes $N$ D3-brane system such that each brane first lying 
in the $(y^2, y^4, \zeta)$ space rotates  by an angle $\theta_{\alpha}$ ($\alpha=1, \cdots, N$) in the 
$y^1$-$y^2$ and $y^3$-$y^4$ planes  as 
$(\xi, \eta)\rightarrow (\e^{i\theta_{\alpha}}\xi, \e^{-i\theta_{\alpha}}\eta)$ 
where $\xi=y^1+iy^2$ and $\eta=y^3+iy^4$,
which rotation belongs to SU(2) group.

Since we apply the T-duality map from 
the type IIA to the type IIB theory along $\zeta=z^5$\,, 
the smeared-out solution yields
\Eqrsubl{D3:fields:Eq}{
F_{(5)}&=&\left(1\pm \ast\right)
d\left(h^{-1}\right)dt\wedge dy^5 \wedge 
\left\{\sum_{\alpha=1}^Nf_{(\alpha)}
\left(R_{(\alpha)}\right)^2_idy^i\wedge 
\left(R_{(\alpha)}\right)^4_jdy^j\right.\nn\\
&&\left. ~~~~~~~-\sum_{a<b}^N f_{(\alpha)}f_{(\beta)}
\sin^2\left(\theta_{\alpha}-\theta_{\beta}\right)
\left(dy^1\wedge dy^3-dy^2\wedge dy^4\right)\right\},\\
\e^{2\phi}&=&\e^{2\phi_0}\,, 
}
where $\phi_0$ is constant and $\ast$ denotes the Hodge dual operator 
in the ten-dimensional background. 
Using the above results, the field equations give 
\Eqrsubl{D3:eq:Eq}{
&&R_{ab}(\Zsp)=0\,,
   \label{D3:Ricci:Eq}\\
&&f_{(\alpha)}(t, z)=\tilde{f}_{(\alpha)}(t)+\bar{f}_{(\alpha)}(z)\,,
   \label{D3:h:Eq}\\
&&\pd_t^2\tilde{f}_{(\alpha)}=0\,,~~~~
\pd_t\tilde{f}_{(\alpha)}\pd_t \tilde{f}_{(\beta)}=0~(\alpha\ne\beta)\,,~~~~
\lap_{\Zsp}\bar{f}_{(\alpha)}=0\,.
   \label{D3:h2:Eq}
}

This gives the exact solution for $N$ D3-brane system with different angles
for giving  Ricci flat $\Zsp$ space.
 As D2-brane system,  only one brane can be time-dependent, 
 \begin{eqnarray}
 \tilde{f}_{(\alpha)}=c_{(\alpha) 0} t+c_{(\alpha)1}
 \,,
 \end{eqnarray}
 where $c_{(\alpha) 0}, c_{(\alpha)1}$ are constant,
 and 
 the other $\tilde{f}_{(\beta)}=c_{(\beta)1}$ ($\beta\neq \alpha$) are constant.

For the metric $u_{ab}=\delta_{ab}$, where $\delta_{ab}$ are 
the four-dimensional flat Euclidean metric,
we find the exact solution 
\Eq{
\bar{f}_{(\alpha)}(z)=\bar c_{\alpha}+\frac{Q_{\alpha}}{|z^a-z^a_{(\alpha)}|^{2-d_\Zsp}},
}
where 
$\bar c_{\alpha}$, $Q_{\alpha}$ and $z^a_{(\alpha)}$ are 
constant parameters and  and 
  $d_\Zsp  (=0,\cdots, 3) $ is the smeared dimensions in the $\Zsp$ space.
  In the case of $d_\Zsp=2 $,
  the power function must be replaced by $\ln |z^a-z^a_{(\alpha)}|$.
 $Q_{\alpha}$ denotes a charge (or mass) of D3-brane and
$z^a=z^a_{(\alpha)}$ is the position of D3-brane.

\subsection{D4-brane with angle}
\label{sec:D4}
Now we will give the dynamical solution of  
$N$ D4-brane system after we apply a T-duality in the $z^4$ direction of the 
ten-dimensional spacetime (\ref{D3:metric:Eq}).
The T-duality relations from type IIB to type IIA are given by
\cite{Bergshoeff:1994cb, Bergshoeff:1995as, 
Breckenridge:1996tt, Costa:1996zd}
\Eqr{
&&g^{(\rm A)}_{\zeta'\zeta'}=\frac{1}{g^{(\rm B)}_{\zeta'\zeta'}}\,,~~~~
g^{(\rm A)}_{MN}=g^{(\rm B)}_{MN}
-\frac{g^{(\rm B)}_{\zeta' M}g^{(\rm B)}_{\zeta' N}
-B^{(\rm B)}_{\zeta' M}B^{(\rm B)}_{\zeta' N}}
{g^{(\rm B)}_{\zeta'\zeta'}}\,,~~~~
g^{(\rm A)}_{\zeta' M}=-\frac{B^{(\rm B)}_{\zeta' M}}{g^{(\rm B)}_{\zeta'\zeta'}}\,,\nn\\
&&\e^{2\phi_{(\rm A)}}=\frac{\e^{2\phi_{(\rm B)}}}{g^{(\rm B)}_{\zeta'\zeta'}}\,,~~~~
C_{M}=C_{\zeta' M}+\chi B_{\zeta' M}^{(\rm B)}\,,~~~~C_\zeta'=-\chi\,,\nn\\
&&B^{(\rm A)}_{MN}=B^{(\rm B)}_{MN}
+2\frac{B^{(\rm B)}_{\zeta' [M}\,g^{(\rm B)}_{N]\zeta'}}{g^{(\rm B)}_{\zeta'\zeta'}}\,,~~~~
B^{(\rm A)}_{\zeta' M}=-\frac{g^{(\rm B)}_{\zeta' M}}{g^{(\rm B)}_{\zeta'\zeta'}}\,,~~~~
C_{\zeta' MN}=C_{MN}
+2\frac{C_{\zeta' [M}\,g^{(\rm B)}_{N]\zeta'}}{g^{(\rm B)}_{\zeta'\zeta'}}\,,\nn\\
&&C_{MNP}=C_{MNP\zeta'}+\frac{3}{2}\left(
C_{\zeta'[M}\,B^{(\rm B)}_{NP]}
-B^{(\rm B)}_{\zeta' [M}\,C_{NP]}
-4\frac{B^{(\rm B)}_{\zeta' [M}\,C_{|\zeta'|N}g^{(\rm B)}_{P]\zeta'}}
{g^{(\rm B)}_{\zeta'\zeta'}}\right)\,,
   \label{D4:duality:Eq}
}
where $\zeta'=z^4$ is the coordinate to which the T-duality is performed.  
The indices $M$\,, $N$ and $P$ denote the other coordinates:
0, $y^i (i=1,\cdots,2)$, $\zeta$ and $z^a (a=1,2, 3)$.  
We delocalize the D3-brane solution in the transverse
coordinate $\zeta'=z^4$\,. 
With the relations in (\ref{D4:duality:Eq}), 
applying T-duality along this direction to produce a system of $N$ D4-brane system
with different angles, 
the type IIA metric is given by 
\Eqrsubl{D4:metric:Eq}{
ds^2&=&h^{-3/8}\left[-dt^2+\gamma_{ij}dy^idy^j+d\zeta^2+d{\zeta'}^2
+h u_{ab}(z)dz^adz^b\right],\\
\gamma_{ij}dy^idy^j&=&\delta_{ij}dy^idy^j+
\sum_{\alpha=1}^Nf_{(\alpha)}\left[
\left\{{\left(R_{(\alpha)}\right)^1}_idy^i\right\}^2
+\left\{{\left(R_{(\alpha)}\right)^3}_jdy^j\right\}^2\right],\\
h(t, z)&=&1+f(t, z)\,,
}
where the function $f$ and the rotation matrix $R_{(\alpha)}$ associated 
with the $\alpha$th D4-brane is defined as (\ref{D2:fields:Eq}). 
The other fields are a straightforward generalization of
the case of a static D4-brane system with certain angles 
in the type IIA low energy effective string theory 
\cite{Breckenridge:1997ar, Youm:1997hw}: 
\Eqr{
F_{(4)}&=&dy^i\wedge dy^j\epsilon_{ijk}\,\pd_k
\left[\sum_{\alpha=1}^Nf_{(\alpha)}
\left(R_{(\alpha)}\right)^2_{~\ell} dy^\ell\wedge 
\left(R_{(\alpha)}\right)^4_{~m} dy^m\right],\\
\e^{2\phi}&=&h^{-1/2}. 
}
Then, the field equations reduce to
\Eqrsubl{D4:eq:Eq}{
&&R_{ab}(\Zsp)=0\,,
   \label{D4:Ricci:Eq}\\
&&f_{(\alpha)}(t, z)=\tilde{f}_{(\alpha)}(t)+\bar{f}_{(\alpha)}(z)\,,
   \label{D4:h:Eq}\\
&&\pd_t^2\tilde{f}_{(\alpha)}=0\,,~~~~
\pd_t\tilde{f}_{(\alpha)}\pd_t \tilde{f}_{(\beta)}=0\,,~(\alpha\ne\beta)~~~~
\lap_{\Zsp}\bar{f}_{(\alpha)}=0\,.
   \label{D4:h2:Eq}
} 

This gives the exact solution for $N$ D4-brane system with different angles
for giving  Ricci flat $\Zsp$ space.
 As the other multi brane systems,  only one brane can be time-dependent, 
 \begin{eqnarray}
 \tilde{f}_{(\alpha)}=c_{(\alpha) 0} t+c_{(\alpha)1}
 \,,
 \end{eqnarray}
 where $c_{(\alpha) 0}, c_{(\alpha)1}$ are constant,
 and 
 the other $\tilde{f}_{(\beta)}=c_{(\beta)1}$ ($\beta\neq \alpha$) are constant.

For the metric $u_{ab}=\delta_{ab}$, where $\delta_{ab}$ are 
the three-dimensional flat Euclidean metric,
the harmonic functions $\bar{f}_{\alpha}$ 
is given by 
\Eq{
\bar{f}_{(\alpha)}(z)=\bar c_{\alpha}+\frac{Q_{\alpha}}{|z^a-z^a_{(\alpha)}|^{1-d_\Zsp}},
}
where $\bar c_{\alpha}$, $Q_{\alpha}$ and $z^a_{(\alpha)}$ are constants
and  $d_\Zsp  (=0,\cdots, 2) $ is the smeared dimensions in the $\Zsp$ space.
 In the case of $d_\Zsp=1 $,
  the power function must be replaced by $\ln |z^a-z^a_{(\alpha)}|$.
 $Q_{\alpha}$ denotes the charge (or mass) of D4-brane and  
each D4-brane is located at
$z^a=z^a_{(\alpha)}$.

\subsection{Summary of dynamical D$p$-brane system with angles and their properties}
\label{sec:g}
In order to find the configurations of a higher dimensional D$p$-brane system with angles ($p>5$),
we can repeat the above procedure by use of T-duality.
 For instance, the solution describing angled D5-brane system 
would have a two-dimensional transverse $\Zsp$ space, and then the function 
$\bar{f}_{(\alpha)}(z)$ is given by the harmonic function 
$\ln [z^a-z^a_{(\alpha)}]$ ($d_\Zsp=0$) or $[z^a-z^a_{(\alpha)}]$ ($d_\Zsp=1$),  while the D6-brane system 
would have the appearance of an anisotropic domain wall in one-dimensional transverse $\Zsp$ space, 
and it is not asymptotically flat ($\bar{f}_{(\alpha)}(z)\propto [z^a-z^a_{(\alpha)}]$). 

We just summarize our result for general $N$ 
D$p$-brane system with different SU(2) angles as follows.

\subsubsection{$N$ D$p$-brane system with different SU(2) angles}
The metric is given by 
\Eqrsubl{c:metric:Eq}{
ds^2&=&h^{p-7\over 8}\left[-dt^2+\gamma_{ij}dy^idy^j + \delta_{mn}d\zeta^m d\zeta^n
+h u_{ab}(\Zsp)dz^adz^b\right]\,,
}
with 
\Eqrsubl{c:metric2:Eq}{
\gamma_{ij}dy^idy^j&=&\delta_{ij}dy^idy^j+
\sum_{\alpha=1}^Nf_{(\alpha)}\left[
\left\{{\left(R_{(\alpha)}\right)^1}_idy^i\right\}^2
+\left\{{\left(R_{(\alpha)}\right)^3}_jdy^j\right\}^2\right],\\
h(t, z)&=&1+f(t, z)\,.
}
The number of the dimensions of $\zeta^m$-space is $p-2$, assuming  $ p\geq 2$.
The functions 
$f(t, z)$ and $f_{(\alpha)}(t, z)$ in D$p$-brane 
are given by 
\Eqrsubl{c:h:Eq}{
&&
f(t, z)=\sum_{\alpha=1}^Nf_{(\alpha)}+\sum_{\alpha<\beta}^N
f_{(\alpha)}f_{(\beta)}\sin^2\left(\theta_{\alpha}-\theta_{\beta}\right),
\label{c:a:Eq}\\
&&f_{(\alpha)}(t, z)=\tilde{f}_{(\alpha)}(t)+\bar{f}_{(\alpha)}(z)\,,
\\
&&
\tilde{f}_{(\alpha)}(t)=c_{(\alpha)0} t+c_{(\alpha)1}\,,~~~~~
\bar{f}_{(\alpha)}(z)=\bar c_{\alpha}+\frac{Q_{\alpha}}{|z^a-z^a_{(\alpha)}|
^{5-p-d_\Zsp}}\,,
 \label{c:f:Eq}
}
where $c_{(\alpha)0}$, $c_{(\alpha)1}$, $\bar c_{\alpha}$ are constants, and the rotation matrices $R_{(\alpha)}$
are defined by Eq. (\ref{D2:fields:Eq}).
$d_\Zsp (=0, \cdots, 6-p)$ denotes the number of smeared dimensions in $\Zsp$ space.
 In the case of $d_\Zsp=5-p $,
  the power function must be replaced by $\ln |z^a-z^a_{(\alpha)}|$.
$Q_{\alpha}~ (\alpha=1\,,\cdots\,, N)$ are the charges (or masses) of D$p$-branes and 
each D$p$-brane is  located at $z^a_{(\alpha)}$, respectively. 

\subsubsection{Properties}
One important result with non-trivial angles $(\theta_\alpha\neq 
\theta_\beta)$ is that the non-linear interaction term between branes 
appears as in Eq. (\ref{c:a:Eq}).

From the solution (\ref{c:h:Eq}), we find that 
the worldvolume of the time-dependent brane system 
 in ten dimensions  for the static observer  is given by 
$\sqrt{-g}\propto h^{5(p-3)/8}$.
Hence, if $c_{(\alpha)0}<0$,
 it is contracting  for $p\geq 4$ while 
expanding for $p\leq 2$.
The world volume is constant for the D3 brane system.
On the other hand, the transverse space to 
D$p$-brane always contracts as  $h^{(p+1)/16}$.
For the case with $c_{(\alpha)0}>0$, we find the opposite behavior.

A curvature 
singularity appears at $h=0$ in the 
ten-dimensional spacetime. 
The regular region in ten-dimensional spacetime
is obtained for if and only if $h>0$.
This regular spacetime region is 
bounded by curvature singularities.

If $d_\Zsp\leq 4-p$, although $h$ diverges on the branes, we find a regular spacetime at infinity.
For $d_\Zsp=5-p$\,, since we have 
\Eq{
\bar{f}_{(\alpha)}(z)=Q_{k}\, 
\ln |z^a-z^a_{(\alpha)}|\,,
     \label{c:h3:Eq}
}
 the harmonic function $h$ diverges
 at infinity ($ |z^a-z^a_{(\alpha)}|\rightarrow \infty$) 
 as well as  near D$p$-branes 
($ |z^a-z^a_{(\alpha)}|\approx 0$).
Hence  such a solution may not be physically relevant.
For $d_\Zsp=6-p$, 
the functions $h$ is given by a sum of linear functions of $z$. 
Then this solution gives a regular behavior near branes, although 
the spacetime is not asymptotically flat.

As a result,  the physically relevant solutions 
are classified by their behaviors into 
two classes: $d_\Zsp\le 4-p$ and $d_\Zsp=6-p$. 
The collision of branes shows different behaviors 
as we will see later.

\subsubsection{Compactified spacetime}

Since some dimensions are homogeneous, we can discuss the dynamics of a compactified
spacetime. 
We compactify some dimensions of homogeneous spaces ($\Ysp$-space, $\zeta^m$-space, and $d_\Zsp$ dimensional smeared $\Zsp$-space).
We assume that $\Ysp$-space, $\zeta^m$-space  and 
$d_c(\leq d_\Zsp)$ dimensions in  smeared $\Zsp$-space are compacitified. 
As a result, we find 
$(d+1)$-dimensional inhomogeneous compactified 
spacetime in the Einstein frame as
\Eq{
ds_{d+1}^2=-h^{-{d\over d-1}}dt^2+h^{-{1\over d-1}}dz_d^2
\,,
     \label{d+1:compactified:spacetime}
}
where $d=7-p-d_c$.

In the far region from the branes, since $h\approx c_{(\alpha)0} t$ for $c_{(\alpha)0}>0$, 
this spacetime expands as $|\tau-\tau_0|^{-{1\over d-2}}$  for $d\neq 2$ 
or $\exp[\mp c_{(\alpha)0}(\tau-\tau_0) ]$ for $d=2$, 
where $\tau$ is the cosmic time in the $(d+1)$-dimensional spacetime and 
$\tau_0$ is an integration constant,
which is given by $(\tau-\tau_0) \propto t^{d-2\over 2(d-1)}$  for $d\neq 2$ 
or $(\tau-\tau_0) =\pm  {1\over c_{(\alpha)0}}\ln t$ for $d=2$.
However, since this is an inhomogeneous $(d+1)$-dimensional spacetime, it cannot describe our universe.
It does not describe a time-dependent black hole system either 
because of the present too simple brane configuration\cite{Maeda:2009zi,
Gibbons:2009dr,
Maeda:2009ds}.


In order to discuss cosmology, we must find our three space in the homogeneous spaces 
 ($\Ysp$-space, $\zeta^m$-space, and 
a subset of $d_\Zsp$ dimensional smeared $\Zsp$-space)
by fixing $z^a$ coordinates of unsmeared $\Zsp$-space.
This means that our universe is described by a test 3-brane located at some point 
$z^a=z_{(0)}^a$.
If we choose our three space in $\Ysp$-space, we find our four-dimensional spacetime 
in the Einstein frame as
\Eq{
ds_{4}^2=h^{-{(p+1)d_\Zsp\over 16}}\left(-dt^2+dy_3^2\right)
\,.
     \label{4:compactified:spacetime}
}
Since $h\approx c_{(\alpha)0} t$ ($c_{(\alpha)0} >0 $), the universe expands as
$|\tau-\tau_0|^{-{(p+1)d_\Zsp\over 32-(p+1)d_\Zsp}}$ where $\tau$ is the cosmic time of our universe.
For D3-branes ($p=3$), since $d_\Zsp\leq 4$, the power exponent is always negative,
which shows a power-law inflation when $\tau\rightarrow \tau_0$\cite{power-law_inflation}, 
except for the case of $d_\Zsp=0$ which is just a Minkowski spacetime.

In what follows, using the solutions obtained in this paper, 
we shall discuss the collision of D3-branes with different angles
in ten-dimensional spacetime.

\section{Collision of dynamical D3-brane}
\label{sec:c}
\subsection{Collision of two D3-branes}
We now discuss collision of two D3-branes with different angles. 
The behavior of the harmonic function $\bar{f}_{(\alpha)}(z)$ 
is classified into two classes depending on the number of smeared 
dimensions of the D3-brane, that is, $d_\Zsp\leq 1$ and $d_\Zsp=3$, 
which we will discuss below separately. For $d_\Zsp=2$, 
the harmonic function $\bar{f}_{(\alpha)}(z)$ diverges both at infinity 
and near D3-branes. 
\subsubsection{$d_\Zsp\leq 1$}
The harmonic function $h$ becomes dominant in 
the limit of $z^a\rightarrow \,z^a_{(\alpha)}$ ($\alpha=1$ and 2).
Hence, we recover a static spacetime of the D$p$-brane system
near branes.
On the other hand, 
the function $h$ depends only on time $t$ 
in the limit of $|z^a|\rightarrow \infty$. 
As a result, in the far region from branes, 
the homogeneous spacetime is found.

In order to analyze the brane collision, 
we consider a concrete example as follows:
Two D3-branes are located at 
$z^a_{(1)}=(z_{(1)}, 0, \ldots, 0)$ and $z^a_{(2)}=(z_{(2)}, 0, \ldots, 0)$. 
We assume that $\tilde f_{(1)}$ is time-dependent, and
 we discuss the time evolution separately with respect 
to the signature of a constant $c_{(1)0}$\,, because the behavior 
of spacetime strongly depends on it. Since the spacetime is singular 
at $h(t, z)=0$, the regular spacetime is obtained inside the spacetime region 
restricted by
\Eq{
h(t, z) = 1+f(t, z)>0\,,
}
where the function $f(t, z)$ is given by 
with \Eq{
f(t, z)=
c_{(1) 0}t+c_{(1)1}+c_{(2)1}+{Q_1\over |z^a-z^a_{(1)}|^{2-d_\Zsp}}
+{Q_2\over |z^a-z^a_{(2)}|^{2-d_\Zsp}}\,.
}
We have set $\bar{c}_1=\bar{c}_2=0$  without loss of generality.
Since the spacetime cannot be extended beyond this region, 
the regular spacetime with two D3-branes ends on these singular
hypersurfaces. 
The solution with $c_{(1)0}>0$ is the time reversal one of $c_{(1)0}<0$, 
because the time dependence appears only in the form of 
$c_{(1)0} t$. In the following, we consider the case with $c_{(1)0}<0$.

For $c_{(1)0}<0$ and the appropriate choice of the constant
$c$, if $Q_1, Q_2>0$, 
the ten-dimensional spacetime is nonsingular at the initial time $(t=0)$
because 
the function $h$ is positive everywhere. 
In the limit of $t\rightarrow -\infty$, 
the ten-dimensional spacetime becomes asymptotically 
a time-dependent uniform background 
except for the cylindrical forms of infinite 
throats near branes ($|z^a-z^a_{(\alpha)}|\approx 0$). 
  
  
   As time evolves $(t>0)$, the singularity appears 
from a far region ($|z-z_{(\alpha)}|\rightarrow \infty$) and 
the singular hypersurface erodes the region with 
the large values of $|z-z_{(\alpha)}|$. As a result, only the region of 
near D3-branes remains regular. A singular hypersurface eventually 
surrounds each D3-brane individually and then the regular
regions near D3-branes splits into two isolated throats.

\begin{figure}[h]
 \begin{center}
\includegraphics[keepaspectratio, scale=0.6]{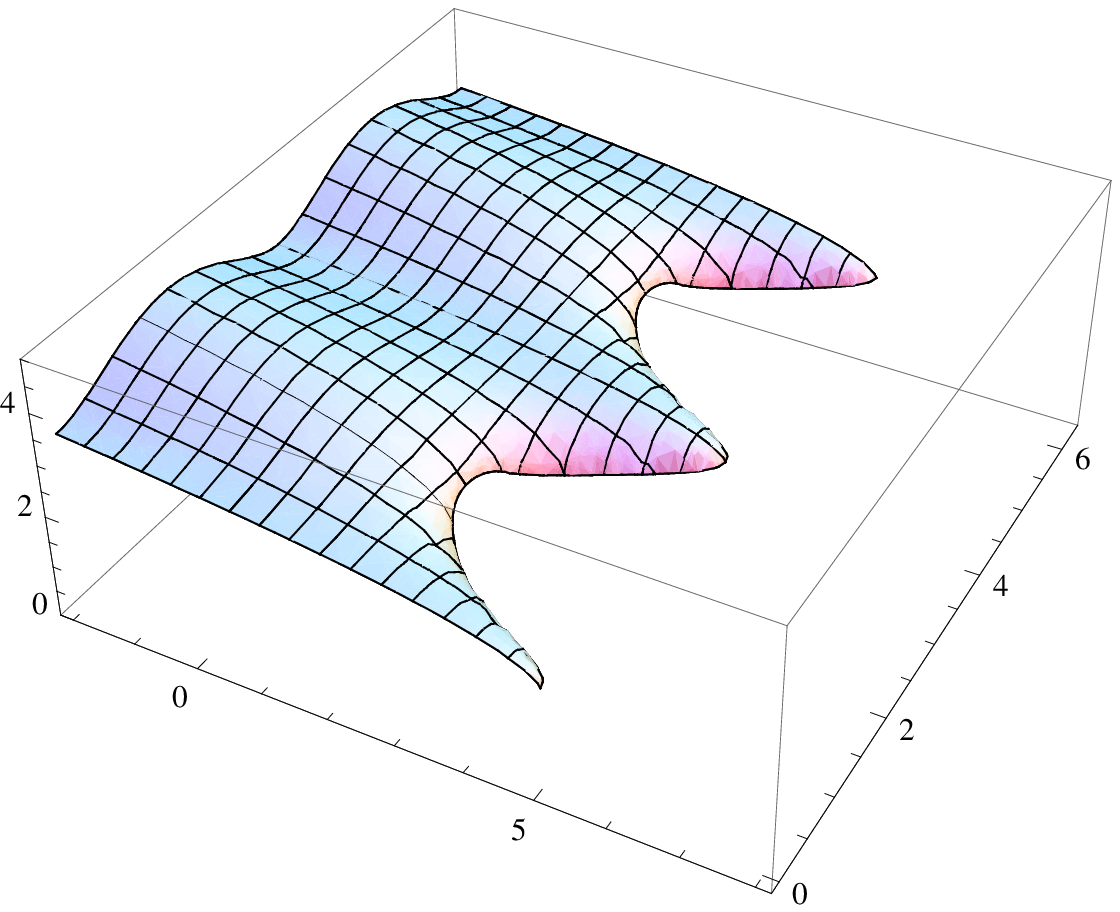}
\put(-205,110){$d(t, \theta)$}
\put(-140,10){$t$}
\put(-10,40){$\theta$}
\hskip 2cm
\includegraphics[keepaspectratio, scale=0.6]{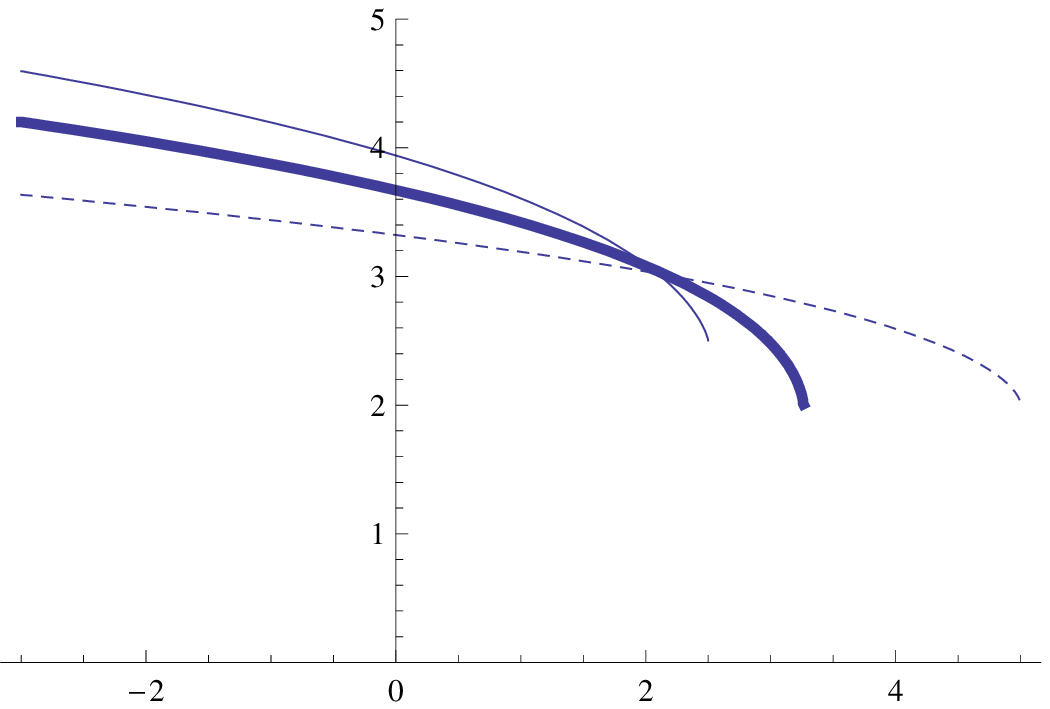}
\put(-125,135){$d(t)$}
\put(10,10){$t$}\\
(a) \hskip 7cm (b)  ~~~~~~
  \caption{\baselineskip 14pt
(a) For the case of $Q_1=Q_2$ in the dynamical D3-branes, 
the proper distance between two dynamical 
 D3-branes given in (\ref{c:distance:Eq}) is depicted. We fix 
$d_\Zsp=1$\,,  $c_{(1)0}=-1$, $Q_1=Q_2=1$, 
 $z_{(1)}=-1\,,~z_{(2)}=1$\,, and $c_{(1)1}=c_{(2)1}=1$ 
for the angle $0\le \theta\le 2\pi$\,. 
A singularity appears between two D3-branes and the spacetime split into
two isolated brane throats before they collide. 
(b) We also show the proper distance $d(t)$ between two dynamical 
 D3-branes for $\theta=0$ (dashed curve), $\theta=\pi/4$ 
 (bold curve) and 
$\theta=\pi/2$ (solid curve) from the bottom in the case of 
 $d_\Zsp=1$\,, 
$Q_1=Q_2=1$, $z_{(1)}=-1\,,~z_{(2)}=1$, and , $c_{(1)0}=-1$ in the 
 dynamical D3-brane system. 
Although the proper distance decreases as $t$ increases, 
the distance is still finite when a singularity appears. 
}
  \label{fig:D3-d1}
 \end{center}
\end{figure}

In Fig.~\ref{fig:D3-d1},  for the case of $d_\Zsp=1$, we show 
the proper distance $d(t, \theta)$
between two branes at $z^a_{(1)}$ and $z^a_{(2)}$,
which  is defined by
\begin{eqnarray}
d(t, \theta)&=&\int_{z_{(1)}}^{z_{(2)}} dz
\left[c_{(1)0}\, t+c_{(1)1}+c_{(2)1}+1
+\frac{Q_{1}}{|z-z_{(1)}|^{2-d_\Zsp}}
+\frac{Q_{2}}{|z-z_{(2)}|^{2-d_\Zsp}}\right.\nn\\
&&\left.
+\left(c_{(1)0}\, t+c_{(1)1}+\frac{Q_{1}}{|z-z_{(1)}|^{2-d_\Zsp}}\right)
\left(c_{(2)1}+\frac{Q_{2}}{|z-z_{(2)}|^{2-d_\Zsp}}\right)\sin^2\theta
\right]^{1/4}\,,
\label{c:distance:Eq}
\end{eqnarray}
where we have defined $z\equiv z^1$, and $d_\Zsp$ denotes the number of 
smeared dimension.
This is a monotonically decreasing function of $t$ if $c_{(1)0}<0$. 
In Fig. ~\ref{fig:D3-d1}, we choose $c_{(1)0}=-1$, $Q_1=Q_2=1$ and $z_{(1)}=-z_{(2)}=-1$.

Fig.~\ref{fig:D3-d1} shows that a singularity appears earlier
as he angle $\theta$ increases from 0 to $\pi/2$,
but the singularity always appears 
before the distance $d$ vanishes. 
Then, a singularity between two D3-branes
forms before they collide each other.

There is no qualitative difference 
when we have the different magnitudes of charges. 
Two branes approach very slowly, but
a singularity eventually appears at a finite distance 
and the spacetime splits into two isolated D3-brane throats. 

We should also mention about the case with different signs of charges (a system of the D3-brane and anti D3-brane). 
If $Q_1$ and $Q_2$ have the different signs, we find 
that there always exists a singularity between two branes.
Hence we cannot even set up such a situation from the beginning.
It is, however,  not the case for $d_\Zsp=3$
(See the next subsection).

Hence, we cannot
discuss a D3-brane collision in the case of $d_\Zsp\leq 1$.

\subsubsection{$d_\Zsp=3$}

Next we consider the case $d_\Zsp=3$\,. 
We set two D3-branes with a brane charge $Q_1$ at $z=z_{(1)}$ and 
the other $Q_2$ at $z=z_{(2)}$. The solution for 
$\bar{f}_{(\alpha)}(z)$ is obtained explicitly as
\Eq{
\bar{f}_{(\alpha)}(z)=Q_{k}\,  |z-z_{(\alpha)}|\,,
}
where $Q_{k}$
 and $z_{(\alpha)}$ are constant parameters. We set 
 $\bar c_{\alpha}=0$ without loss of generality.

The proper distance between two D3-branes is given by
\begin{eqnarray}
d(t, \theta)&=&\int_{z_{(1)}}^{z_{(2)}} dz
\left[c_{(1)0}\, t+c_{(1)1}+c_{(2)1}+1
+Q_{1}|z-z_{(1)}|+Q_{2}|z-z_{(2)}|\right.\nn\\
&&\left.
+\left(c_{(1)0}\, t+c_{(1)1}+Q_{1}|z-z_{(1)}|\right)
\left(c_{(2)1}+Q_{2}|z-z_{(2)}|\right)\sin^2\theta
\right]^{1/4}.
\label{c:distance2:Eq}
\end{eqnarray}

\begin{figure}[h]
 \begin{center}
          \includegraphics[keepaspectratio, scale=0.6]{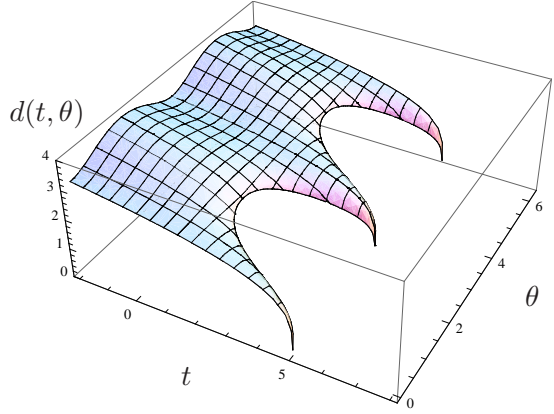}
\put(-205,110){$d(t, \theta)$}
\put(-140,10){$t$}
\put(-10,40){$\theta$}\\
\caption{\baselineskip 14pt
For the case of $Q_1=Q_2$ in the dynamical D3-branes, 
the proper distance between two dynamical 
 D3-branes given in (\ref{c:distance:Eq}) is depicted. We fix 
 $d_\Zsp=3$\,, 
 $c_{(1)0}=-1$, $Q_1=Q_2=1$, 
 $z_{(1)}=-1\,,~z_{(2)}=1$\,, and $c_{(1)1}=c_{(2)1}=1$ 
 for the $0\le \theta\le 2\pi$\,.  
The distance decreases, and
then two D3-branes collide each other. 
The proper distance rapidly 
vanishes near two branes collide for the case of 
$\theta=0\,,~\theta=\pi\,,~\theta=2\pi$\,.
}
  \label{fig:D3-1}
 \end{center}
\end{figure}

\begin{figure}[h]
 \begin{center}
\includegraphics[keepaspectratio, scale=0.6]{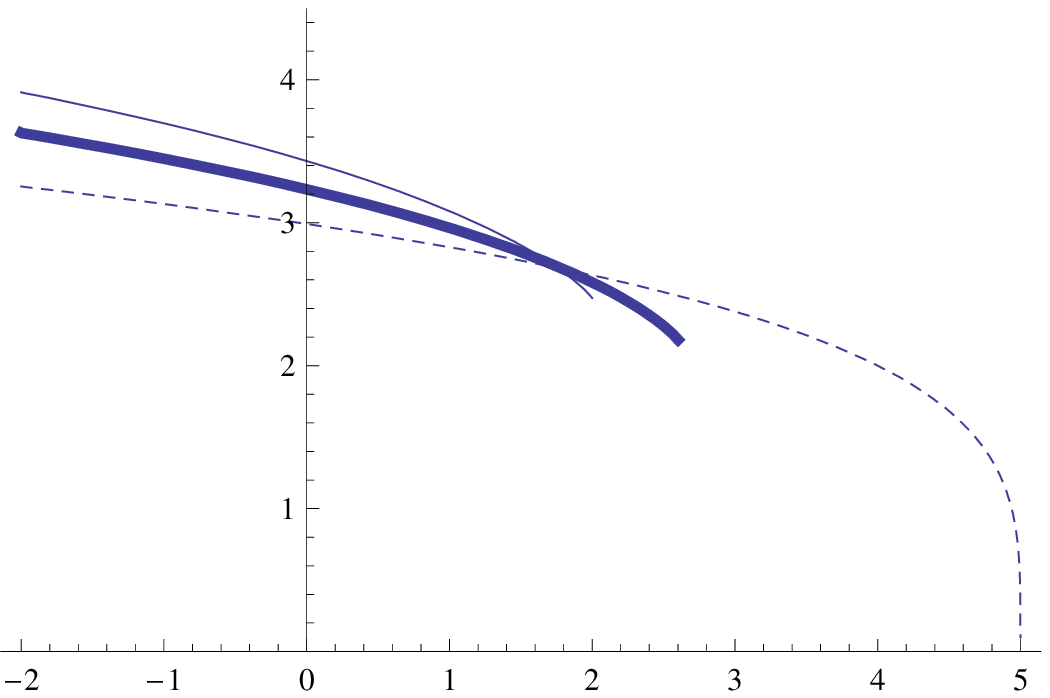}
\put(-140,135){$d(t)$}
\put(10,10){$t$}
\hskip 2cm
\includegraphics[keepaspectratio, scale=0.6]{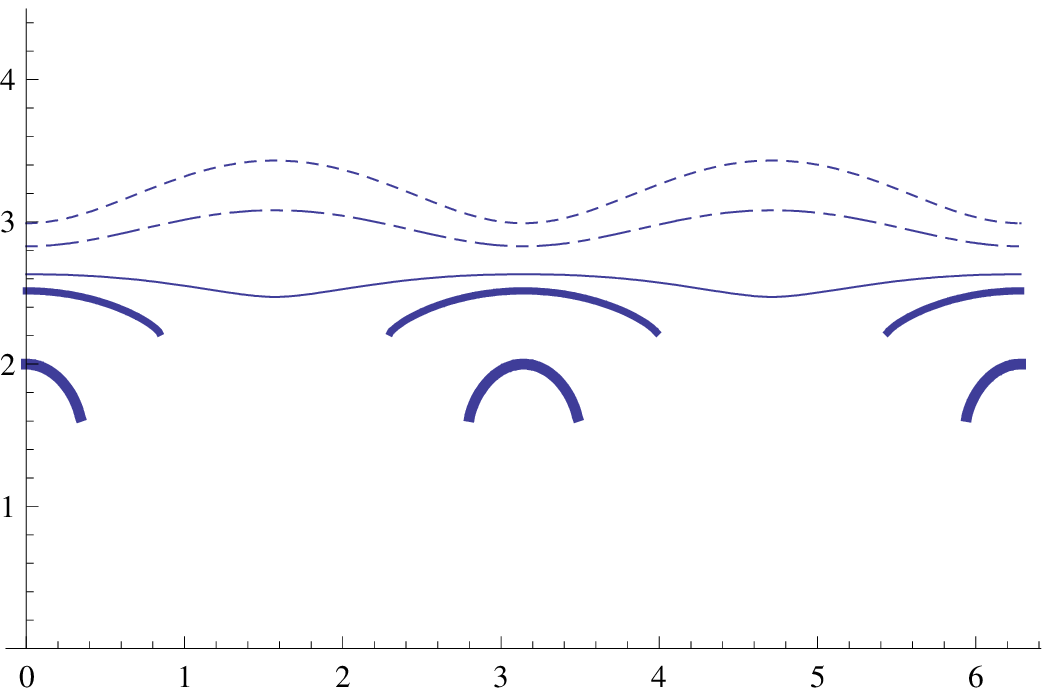}
\put(-190,135){$d(\theta)$}
\put(10,10){$\theta$}\\
(a) \hskip 7cm (b)  ~~~~~~
  \caption{\baselineskip 14pt
(a) We show the proper distance between two dynamical 
 D3-branes for $\theta=0$ (dashed curve), $\theta=\pi/4$ 
 (bold curve) and 
$\theta=\pi/2$ (solid curve) from the bottom in the case of 
 $d_\Zsp=3$\,, 
$Q_1=Q_2=1$, $z_{(1)}=-1\,,~z_{(2)}=1$, and $c_{(1)0}=-1$ in the 
 dynamical D3-brane system. 
 If we set $\theta=0$, it causes the complete collision at $t=5$ 
 simultaneously.
For the case of $\theta\ne n\pi~(n=\,$integer), it
is still finite when a singularity appears. 
(b) We also show the snapshots at $t=0$ (dashed line), 
$t=1$ (dashed-dotted line), $t=2$ (solid line), $t=2.5$ (bold line), 
and $t=4$ (thick bold line). The proper distance 
$d(t\,,~\theta)$ vanishes at $t=5$\,, $\theta= n\pi~(n=\,$integer)\,.
}
  \label{fig:D3-2}
 \end{center}
\end{figure}

The proper distance is a monotonically decreasing function of $t$ 
if we set again $c_{(1)0}<0$\,.
We illustrate $d(t, \theta)$  
for the case of the D3-brane system in Fig.~\ref{fig:D3-1}.   
We also depict the proper distance $d(t)$ 
for a given angles $\theta=0,\pi/4$ and $\pi/2$, 
and the angular dependence of the proper distance $d(\theta)$
at some particular times $t=0, 1, 2, 2.5, 4$ and $5$ 
in Fig.~\ref{fig:D3-2}. 
We set $c_{(1)0}=-1$\,, $c_{(1)1}=c_{(2)1}=1$\,, 
$z_{(1)}=-1$\,, $z_{(2)}=1$\,, and $Q_1=Q_2=1$.


The angle dependence is the similar to the case of $d_\Zsp\leq 1$.
The proper distance $d$ between two branes never vanishes.
For the case of the equal charges ($Q_1=Q_2$), however, there is one exceptional case, 
which is  two D3-brane system  with the trivial angle 
($\theta=0$ or $\pi$).
In this case, as we can see in Figs. \ref{fig:D3-1} and \ref{fig:D3-2},
 $d$ vanishes just  when a singularity appears.
Two D3-branes collide completely and a singularity appears at
the same time. 
Hence we conclude that 
two branes will never collide each other except the case 
with the trivial angles. 

\begin{figure}[h]
 \begin{center}
\includegraphics[keepaspectratio, scale=0.6]{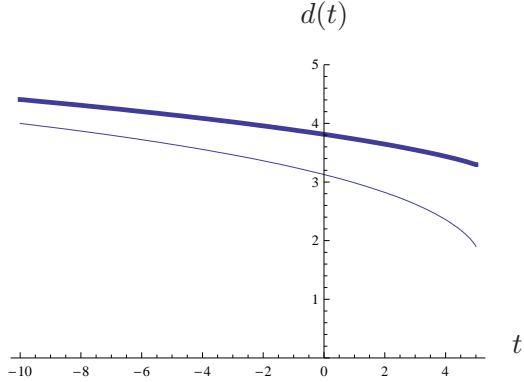}
\put(-70,135){$d(t)$}
\put(10,10){$t$}\\
  \caption{\baselineskip 14pt 
We  depict the proper distance between two dynamical 
 D3-branes for the cases of 
 $10Q_1=Q_2=10$ (thick bold line) and $2Q_1=Q_2=2$ (solid curve).
  We choose  
$\theta=0$\,, $d_\Zsp=3$, $z_{(1)}=-1\,,~z_{(2)}=1$, 
and  $c_{(1)0}=-1$. 
All proper distances are still finite when a singularity appears. 
}
  \label{fig:D3-diff_charge}
 \end{center}
\end{figure}

How about the different values of charges ?
We then show the case with different magnitudes of charges 
in Fig. \ref{fig:D3-diff_charge}.
In this figure, we have just shown the case with $\theta=0$,
which may give the closest distance when a singularity appears.
This figure shows that any two-brane system with the different magnitudes of charges never collide 
when a singularity appears.

\begin{figure}[h]
 \begin{center}
          \includegraphics[keepaspectratio, scale=0.6]{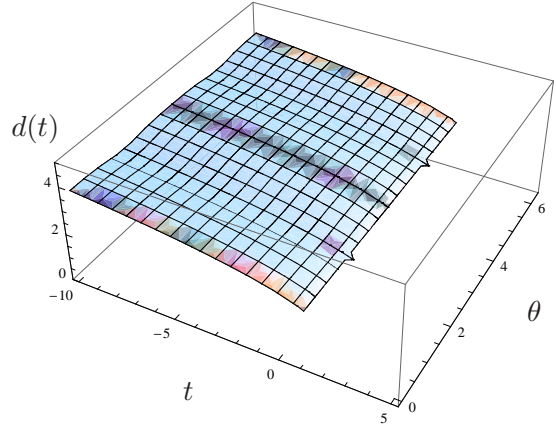}
\put(-205,110){$d(t)$}
\put(-140,10){$t$}
\put(-10,40){$\theta$}\\
   \caption{\baselineskip 14pt
 For the case of $Q_1=-Q_2$ in the dynamical D3-branes, 
the proper distance between two dynamical 
 D3-branes given in (\ref{c:distance:Eq}) is depicted. We fix 
 $d_\Zsp=3$\,,  $c_{(1)0}=-1$, $Q_1=-Q_2=1$, 
 $z_{(1)}=-1\,,~z_{(2)}=1$\,, and $c_{(1)1}=c_{(2)1}=1$ 
 for the $0\le \theta\le 2\pi$\,.  
In this case, a singularity
appears at $t<0$ when the distance is still finite. 
Then, the solution does not describe the collision of two D3-branes.
}
  \label{fig:D3-3}
 \end{center}
\end{figure}

\begin{figure}[h]
 \begin{center}
\includegraphics[keepaspectratio, scale=0.6]{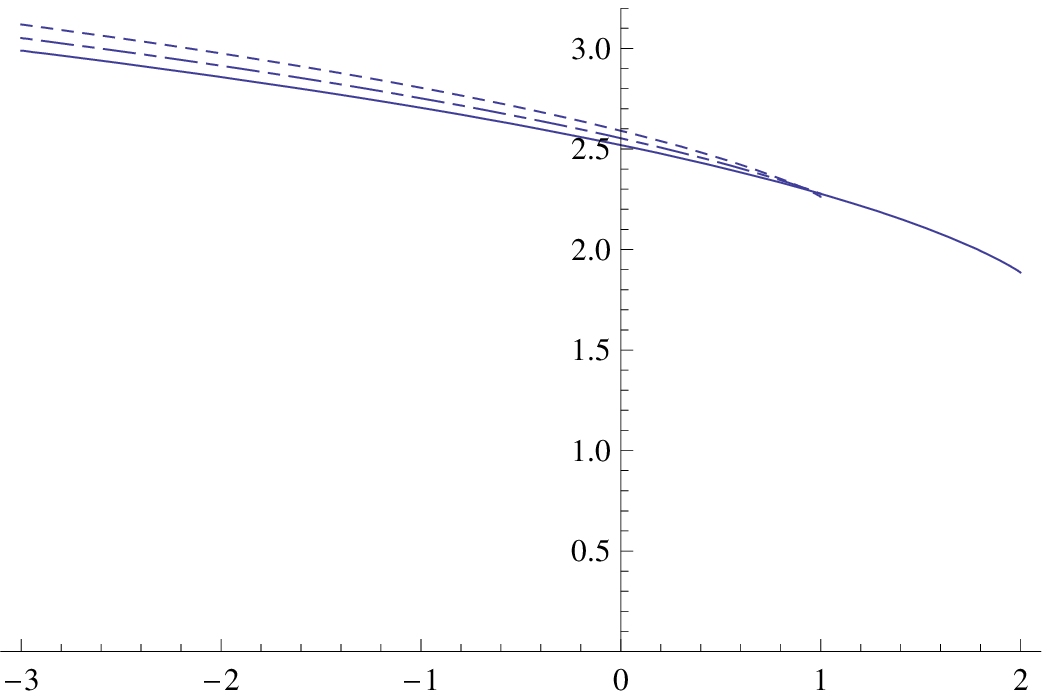}
\put(-85,135){$d(t)$}
\put(10,10){$t$}
\hskip 2cm
\includegraphics[keepaspectratio, scale=0.6]{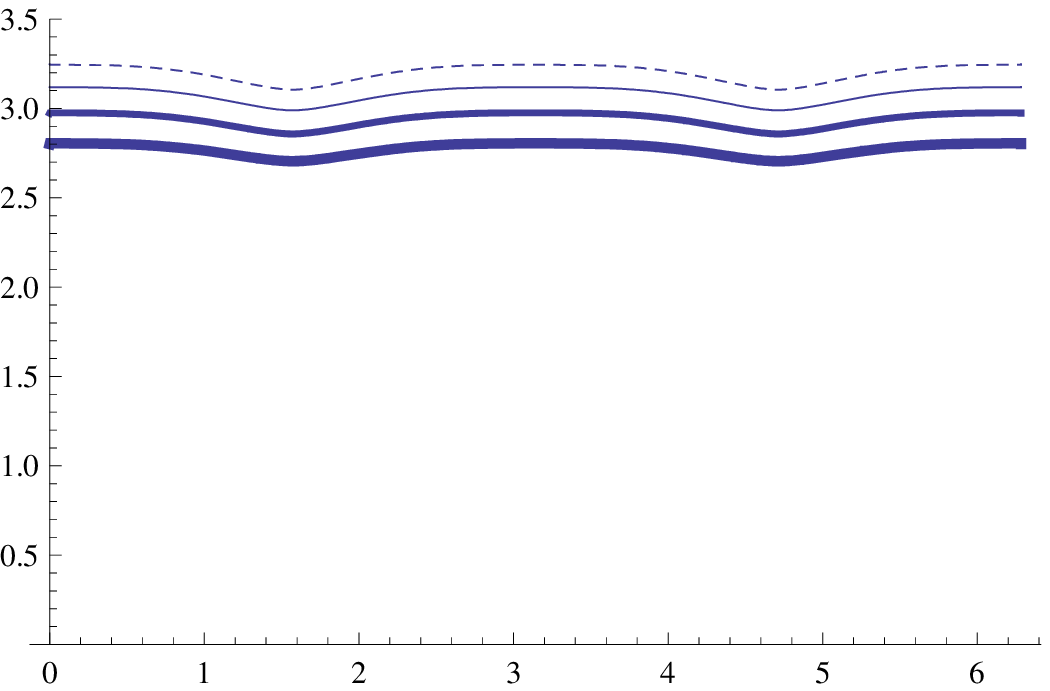}
\put(-190,135){$d(\theta)$}
\put(10,10){$\theta$}\\
(a) \hskip 7cm (b) ~~~~~~
  \caption{\baselineskip 14pt
(a) 
We show the proper distance at $\theta=0$ (dotted line), 
$0.35\pi$ (dashed-dotted line), $0.5\pi$ (solid line) 
in the case of dynamical D3-brane system. 
We set $d_\Zsp=3$\,, 
$Q_1=-Q_2=1$, $z_{(1)}=-1\,,~z_{(2)}=1$, and $c_{(1)0}=-1$. 
Although the proper distance decreases as $t$ increases, 
the distance is still finite when a singularity appears at 
$t<0$\,. 
(b)We also depict the snapshots at $t=-4$ (dashed line), 
$t=-3$ (solid line), $t=-2$ (bold line), and $t=-1$ (thick bold line). 
Although the proper distance decreases as t increases, 
the distance is still finite when a singularity appears at $t<0$\,. 
}
  \label{fig:D3-4}
 \end{center}
\end{figure}

We also show the proper distance between two branes
 for the case with different signs of charges 
such that  $Q_1=1$ and $Q_2=-1~$ in 
Figs.~\ref{fig:D3-3} and \ref{fig:D3-4}\,. 
In this case, it describes the system of the D3-brane 
and the anti D3-brane with the same masses of $M_1=M_2=1$.


\begin{figure}[h]
 \begin{center}
\includegraphics[keepaspectratio, scale=0.6]{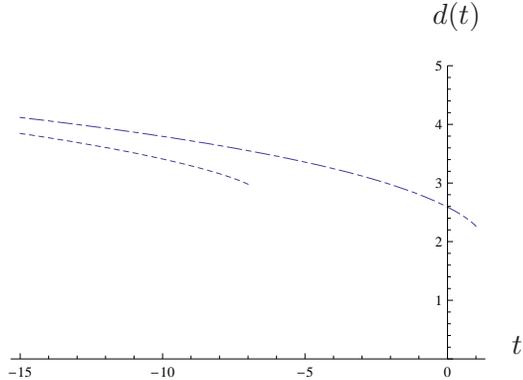}
\put(-20,135){$d(t)$}
\put(10,10){$t$}\\
  \caption{\baselineskip 14pt
We depict the proper distance between two dynamical 
 branes for the cases of 
 $Q_1=-Q_2=1$ (dashed-dotted line) and $5Q_1=-Q_2=5$ 
 (dashed curve) . We choose  
$\theta=0$\,, $d_\Zsp=3$, $z_{(1)}=-1\,,~z_{(2)}=1$, 
and  $c_{(1)0}=-1$. 
All proper distances are still finite when a singularity appears. 
}
  \label{fig:D3-diff_charge2}
 \end{center}
\end{figure}

Although two branes approach each other,  a 
singularity always appears before two branes collide. 
It is different from the case with the same charges ($Q_1=Q_2$), 
in which two branes can collide when $\theta =0$ or $\pi$.
The proper distance $d(t, \theta)$ never vanishes 
for the present case. 
This result does not change even if the magnitudes of two charges are
different, as we show in Fig. \ref{fig:D3-diff_charge2}.

We then conclude that for the case of $d_\Zsp=3$,
only two-brane system with the same charges as well as with 
the trivial angle ($\theta=0$ or $\pi$) gives a complete 
brane collision.
\section{Discussions}
  \label{sec:d}

In the present paper, we have discussed the dynamical D$p$-brane solution 
which describes several D$p$-branes
oriented at angles with respect to one another. 
Since the corresponding background field configurations remain largely 
unexplored, we have presented one such class of solutions 
in the time-dependent D$p$-brane background. 
The dynamical solution which we have obtained in this paper describes any
number of D$p$-brane whose relative orientations are given by certain 
SU(2) rotations. 
These are functions of time that becomes static near D-branes 
with the simplest possible dependence on the warp factor. 
In the far region from angled D$p$-brane in the ten-dimensional 
background, the solutions give purely contracting or expanding 
uniform universe. 
The worldvolume of the time-dependent brane 
is also either contracting or expanding in ten dimensions depending 
on the brane dimensions as well as the sign of $c_{(\alpha)0}$ except for 
D$3$-brane system.  The transverse space is always contracting if 
$c_{(\alpha)0}<0$.

We have also discussed the dynamics of angled D3-branes 
models with applications to  
collision of branes. 
The two D3-branes approach at a angle $\theta$ in the direction of 
spatial part of world volume coordinates $y^i$. 
We have studied the dynamics of D3-branes which 
have been smeared along the transverse space to D3-branes. 
If two D3-brane charges are different from each other ($Q_1=-Q_2$), 
the distance between two D3-branes is still finite when a singularity 
appears. Thus, we cannot describe the collision of two D3-branes 
in terms of the solution. 
For $d_\Zsp \le 1$, a singularity again appears 
before D3-branes collide. Then, the topology of the spacetime eventually 
changes so that branes are separated by singular hypersurfaces 
surrounding each D3-brane. 
This behavior also appears if $Q_1=Q_2$ and $d_\Zsp=3$.
A singularity forms at $\theta\ne 0\,,\pi\,,$ 
when the distance is still finite. 
On the other hand, it has been shown that there exists 
a complete collision where the orientation of configurations between 
two smeared D3-branes with $Q_1=Q_2$ are either 0 or $\pi$.
This result may be related to supersymmetry, which may be broken for non-paralell 
branes. 

Although the examples presented in the present paper cannot provide 
a realistic cosmological model, the solution may be utilized to construct 
a cosmological solution  just by introduction of a test brane universe
in higher-dimensions.
We may also construct new type of time-dependent black hole solution 
with non-trivial angles by setting up more complicated brane configuration.
Those subjects are left for a future works.

\section*{Acknowledgments}
This work is supported by Grants-in-Aid from the 
Scientific Research Fund of the Japan Society for the Promotion of Science 
(No. 25400276). 





\end{document}